\documentclass[aps,prb,superscriptaddress,showpacs,floatfix,twocolumn]{revtex4}

\usepackage{graphicx}

\begin{document}
\title{Decoherence in Josephson-junction qubits due to critical current
  fluctuations }

\author{D.J. Van Harlingen}
\affiliation{Department of Physics, University of Illinois at
Urbana-Champaign, Urbana, IL 61801}

\author{T.L. Robertson}
\affiliation{Department of Physics, University of California,
%and Materials Sciences Division, Lawrence Berkeley National
Berkeley, CA 94720 }

\author{B.L.T. Plourde}
\affiliation{Department of Physics, University of California,
%and Materials Sciences Division, Lawrence Berkeley National
%Laboratory,
Berkeley, CA 94720 }

\author{P.A. Reichardt}
\affiliation{Department of Physics, University of California,
%and Materials Sciences Division, Lawrence Berkeley National
%Laboratory,
Berkeley, CA 94720 }

\author{T.A. Crane}
\affiliation{Department of Physics, University of Illinois at
Urbana-Champaign, Urbana, IL 61801}

\author{John Clarke}
\affiliation{Department of Physics, University of California,
Berkeley, CA 94720 }

\pacs{85.25.Cp, 85.25.Am, 03.67.Lx}

\date{12 December 2003}

\begin{abstract}

  We compute the decoherence caused by $1/f$ fluctuations at low
  frequency $f$ in the critical current $I_0$ of Josephson junctions
  incorporated into flux, phase, charge and hybrid flux-charge
  superconducting quantum bits (qubits). The dephasing time
  $\tau_{\phi}$ scales as $I_0/ \Omega \Lambda S_{I_0}^{1/2}(1$ Hz$)$,
  where $\Omega / 2\pi$ is the energy level splitting frequency,
  $S_{I_0}(1$ Hz$)$ is the spectral density of the critical current noise
  at $1$ Hz, and $\Lambda \equiv |I_0 d \Omega / \Omega d I_0|$ is a
  parameter computed for given parameters for each type of qubit that
  specifies the sensitivity of the level splitting to critical current
  fluctuations.  Computer simulations show that the envelope of the
  coherent oscillations of any qubit after time $t$ scales as $\exp
  (-t^2/ 2 \tau_{\phi}^2)$ when the dephasing due to critical current
  noise dominates the dephasing from all sources of dissipation.  We
  compile published results for fluctuations in the critical current
  of Josephson tunnel junctions fabricated with different technologies
  and a wide range in $I_0$ and $A$, and show that their values of
  $S_{I_0}(1$ Hz$)$ scale to within a factor of three of $\left[ 144
    \left( I_0/\mu{\rm A}\right)^2/ \left( A/ \mu{\rm m}^2
    \right)\right]($pA$)^2/$Hz at $4.2$ K. We empirically extrapolate
  $S_{I_0}^{1/2}(1$ Hz$)$ to lower temperatures using a scaling
  $T($K$)/4.2$. Using this result, we find that the predicted values of
  $\tau_{\phi}$ at $100$ mK range from $0.8$ to $12$ $\mu$s, and are
  usually substantially longer than values measured experimentally at lower
  temperatures.

\end{abstract}

\maketitle

\section {INTRODUCTION}

Superconducting devices involving Josephson junctions are leading
candidates for quantum bits (qubits) because of their
manufacturability, controllability and scalability. Broadly speaking,
there are three types of superconducting qubits. The first type is the
flux qubit, which consists of a superconducting loop interrupted by
either one \cite{Lukens,Leggett} or three \cite{Mooij,Chiorescu}
junctions. When the qubit is biased at the degeneracy point the two
states represented by magnetic flux pointing up and pointing down are
superposed to produce symmetric and antisymmetric eigenstates. Quantum
coherent behavior has been verified by means of spectroscopic
measurements of the level splitting of these states
\cite{Lukens,Mooij} and by the observation of Rabi
oscillations. \cite{Chiorescu}
The second type of qubit is based on the charge
degree of freedom, and consists of a nanoscale superconducting island
coupled to a superconducting reservoir via a Josephson junction. The
two quantum states differ by a single Cooper pair. Superpositions of
these states have been demonstrated through Rabi oscillations,
\cite{Nakamura1} and signatures of the entanglement of two charge
qubits have been observed. \cite{Pashkin}  These two qubit types are
distinguished by whether the Josephson coupling energy $E_J$ or the
charging energy $E_C$ dominates the junction dynamics.  A hybrid
charge-flux device was operated in the crossover between these two
regimes, at its degeneracy points in both charge and
flux; \cite{Quantronium,Cottet} it exhibited the longest dephasing time
yet reported for a superconducting qubit, about $0.5$ $\mu$s.  The third
type is the phase qubit, which consists of a single Josephson junction
current-biased in the zero voltage state. \cite{Martinis-Rabi,Han}  In
this case, the two quantum states are the ground and first-excited
states of the tilted potential well, between which Rabi oscillations
have been observed. Unlike the other qubits, the phase qubit does not
have a degeneracy point.

For all these qubits, the measured decoherence times are
substantially shorter than predicted by the simplest models of
decoherence from dissipative sources and than would be necessary
for the operation of a quantum computer. As a result, there is an
ongoing search to identify additional sources of dephasing. In the
case of charge qubits, the coherence times have been limited by
low frequency fluctuations of background charges in the substrate
which couple capacitively to the island, thus dephasing the
quantum state. \cite{Nakamura2}  Flux and phase qubits are
essentially immune to fluctuations of charge in the substrate,
and, by careful design and shielding, can also be made insensitive
to flux noise generated by either the motion of vortices in the
superconducting films or by external magnetic noise. The
flux-charge hybrid, operated at its double degeneracy point, is
intrinsically immune to both charge and flux fluctuations.
However, all of these qubits remain sensitive to fluctuations in
the critical current of the tunnel junctions at low frequency $f$,
which lead to variations in the level splitting frequency over the
course of the measurement and hence to dephasing.

Martinis {\it et al.} \cite{Martinis-noise} analyzed decoherence
in phase qubits due to low frequency critical current
fluctuations, and Paladino {\it et al.} \cite{Paladino} treated
decoherence in charge qubits due to low frequency charge noise. In
this paper, we explore the effects of low frequency noise in the
critical current on the dephasing times $\tau_\phi$ in various
superconducting qubits incorporating Josephson junctions, and
compare our results with measured decoherence times.  In Sec. II
we discuss two sources of low frequency fluctuations in
superconducting circuits and explain how they induce dephasing. In
Sec. III we calculate the sensitivity of several Josephson qubit
schemes to critical current variations, using parameters from
recent experiments reporting dephasing times.  In Sec. IV we
compile a list of measurements of the critical current noise in a
variety of junctions and obtain a ``universal value'' that we use
in subsequent estimates of decoherence times.  In Sec. V we
estimate dephasing times limited by $1/f$ noise, using numerical
simulations to elucidate the dephasing process.  Section VI
contains some concluding remarks.

\section {DECOHERENCE MECHANISM FOR LOW FREQUENCY NOISE}

We consider two intrinsic sources of low frequency noise in
superconducting devices which can cause dephasing. Flux vortices
hopping between pinning sites in superconducting films, illustrated in
Fig. \ref{1-f decoherence mechanism}(a), result in fluctuations of
the magnetic flux in multiply-connected superconducting circuits.
Specifically, in superconducting flux qubits operating at the
degeneracy of the left and right circulating current states, external
magnetic flux $\Phi_x$ breaks the degeneracy, causing a second-order
change in the tunneling frequency. This mechanism can usually be made
negligible in devices fabricated with linewidths less than
approximately $(\Phi_0/B)^{1/2}$ for which vortex trapping in the line
is suppressed; \cite{Dantsker96,Clem} here $\Phi_0\equiv h/2e$ is the flux
quantum and $B$ is the field in which the device is cooled.

A more serious problem is critical current fluctuations caused by
charge trapping at defect sites in the tunneling barrier, as in Fig.
\ref{1-f decoherence mechanism}(b).  In the prevailing picture,
trapped charges block tunneling through a region of the junction due
to the Coulomb repulsion, effectively modulating the junction area. In
general, a single charge fluctuator produces a two-level, telegraph
signal in the critical current of a junction, characterized by
lifetimes in the untrapped (high critical current) state $\tau_u$ and
the trapped (low critical current) state $\tau_t$.  This produces a
Lorentzian bump in the power spectral density with a characteristic
time $\tau_{eff}= (1/\tau_t + 1/\tau_u)^{-1}$. The dynamics of such
fluctuators in junctions have been extensively studied \cite{Wakai1,
  Rogers, Rogers-2}, and the lifetimes have been measured as a
function of temperature and voltage bias.  There is strong evidence
from the voltage dependence that the dominant charges enter the
barrier from one electrode and exit to the other, and that the
fluctuators exhibit a crossover from thermal activation to tunneling
behavior at about $15$ K. In the tunneling regime, the fluctuating
entity has been shown to involve an atomic mass, suggesting that ionic
reconfiguration plays an important role in the tunneling process.
Interactions between traps resulting in multiple level hierarchical
kinetics have been observed, \cite{Wakai2} but usually the traps can
be considered to be local and non-interacting.  In this limit, the
coexisting traps produce a distribution of Lorentzian features that
superimpose to give a $1/f$-like spectrum. \cite{Dutta-Horn,Weissman}

\begin{figure}
 \centering
  \caption{{\it note: figure attached} Effects of low frequency flux and critical current
   fluctuations in a superconducting qubit.  (a) Flux modulation from vortices
   hopping into and out of a loop, and critical current modulation from electrons
   $e^-$ temporarily trapped at defect sites in the junction barrier.
   (b) A single charge trap blocks tunneling over an area $\delta A$,
   reducing the critical current. (c) Fluctuations modify the oscillation frequency,
   inducing phase noise which leads to decoherence in time-averaged ensembles of
   sequential measurements of the qubit observable $Z$.}
 \label{1-f decoherence mechanism}
\end{figure}

The parametric fluctuations in the qubit energy levels introduce
phase noise into the measurement of the probability distribution
of the qubit states.  The key point is that determination of the
qubit state and its evolution with time requires a large number of
measurements. In the presence of low frequency noise, the energy
levels fluctuate during the data acquisition.  This causes an
effective decoherence in the qubit, as illustrated in Fig.
\ref{1-f decoherence mechanism}(c). The resulting decay of the
qubit state probability amplitude reflects the spectrum of the low
frequency noise.

\section {QUBIT SENSITIVITY TO CRITICAL CURRENT FLUCTUATIONS}

We consider a superconducting qubit with quantum states separated
in energy by $\hbar\Omega$, and assume that the splitting depends
on the critical current of one or more Josephson tunnel junctions
in the qubit. The sensitivity of the energy difference to critical
current fluctuations is described by the dimensionless parameter
\begin{eqnarray}
\Lambda &=& \left|I_0 d\Omega/\Omega dI_0\right|,\label{eq:lambda}
\end{eqnarray}
the fractional change in the energy separation for a given fractional
change in the critical current $I_0$.  The value of $\Lambda$ depends
on the qubit architecture, the device parameters, and the bias point.
A large value of $\Lambda$ indicates that a particular qubit type is
vulnerable to decoherence caused by critical current fluctuations;
small values indicate a more robust qubit design for fluctuations of
the same amplitude.  In the following sections, we calculate $\Lambda$ for
a variety of qubit designs and parameters used in recent experiments.
In some cases, we can develop analytical expressions for the energy
separation, which often is a tunneling matrix element, from which
$\Lambda$ can be calculated; in others, it is necessary to carry out
numerical calculations to estimate the response to critical current
changes.

\subsection {One-Junction Flux Qubit (Ground State)}

We first consider the one-junction flux qubit [Fig. \ref{rf SQUID
  ground state analytic}(a)], consisting of a single Josephson
junction of critical current $I_0$ and capacitance $C$ in a loop of
inductance $L$ biased with an applied flux $\Phi_x$.  At the
degeneracy point $\Phi_x = \Phi_0/2$, the energy vs. flux curve is a
degenerate double-well potential given by $V(\phi)=(\Phi_0^2/8\pi^2
L)[2 \beta_L \cos(\phi)+(\phi+\pi+2\pi\Phi_x/\Phi_0)^2]$, in terms of
the junction phase $\phi$. The two states of lowest energy are
approximately symmetric and antisymmetric combinations of localized
states in the left and right wells characterized by clockwise and
counterclockwise circulating currents, between which the ``phase
particle'' tunnels [Fig. \ref{rf SQUID ground state analytic}(b)].
Fluctuations in the flux tilt the potential wells, weakly changing the
tunneling frequency in second order [Fig. \ref{rf SQUID ground state
  analytic}(c)]; however, critical current fluctuations directly
modulate the barrier height, producing an exponential change in the
qubit tunneling frequency [Fig. \ref{rf SQUID ground state
  analytic}(d)].

We now calculate the tunnel splitting, or more precisely the energy
difference between the ground and first excited state, for the
one-junction flux qubit using three different methods.  The purpose of
this pedagogical exercise is to understand in which regimes certain
approximations are valid.  We build on this insight to
analyze other qubits later in this paper.

Our first approach is to approximate the potential with a quartic
polynomial and quote an analytic result for the tunneling
frequency in the semi-classical WKB approximation, \cite{Leggett}
\begin{eqnarray}
\Omega &=& \omega_0 \exp \left[ -\eta(\beta_L-1)^{3/2} \right].\label{eq:rf_wkb_simple}
\end{eqnarray}
Here $\omega_0\equiv2[(\beta_L-1)/LC]^{1/2}$ is the classical
frequency of small oscillations in the bottom of the wells,
$\beta_L\equiv2\pi L I_0/\Phi_0$ is the dimensionless screening
parameter, and $\eta\equiv(8 I_0C\Phi_0^3/\pi^3\hbar^2)^{1/2}$ is a
parameter which describes the ``degree of classicality'' and hence
determines when quantum tunneling is important. \cite{Leggett}  Figure
\ref{rf_SQUID_ground_state numerical}(a) plots $\Omega/2\pi$ {\it vs.}
$\beta_L$ for stated values of $L$ and $C$.

\begin{figure}
\centering
\caption{{\it note: figure attached} One-junction flux qubit.  (a) Schematic. (b) Symmetric double
  well potential for flux bias $\Phi_x=\Phi_0/2$. (c) Flux fluctuation
  $\Delta \Phi$ couples to $\Omega$ only in second order. (d) Critical
  current fluctuation $\Delta I_0$ produces exponential change in
  $\Omega$.}
\label{rf SQUID ground state analytic}
\end{figure}

However, the semi-classical approximation is valid only in the regime
where the bound states in each well nearly form a continuum, which is
far from the case we consider here with only one bound state in each
well.  To obtain the correct splittings for the ground state in the
WKB approximation one must modify Eq. (\ref{eq:rf_wkb_simple}).  A more
accurate result is \cite{garg}
\begin{eqnarray}
\Omega &=& 2 \omega_0 \sqrt{\frac{m \omega_0 \phi_m^2}{\pi \hbar}}
e^A e^{-S_0/\hbar} ,\label{eq:wkb_garg}
\end{eqnarray}
where $S_0$ is the action along the tunneling direction
\begin{eqnarray}
S_0 &=& \int_{-\phi_m}^{\phi_m} \sqrt{2 m V(\phi)} d\phi
,\label{eq:wkb_S0}
\end{eqnarray}
and $A$ is a correction factor
\begin{eqnarray}
A &=& \int_0^{\phi_m} \left[\frac{m \omega_0}{\sqrt{2 m
V(\phi)}}-\frac{1}{\phi_m-\phi}\right]d\phi .\label{eq:wkb_A}
\end{eqnarray}
Here $m=C \left( \Phi_0 / 2 \pi \right)^2$ is the effective mass
of the tunneling particle, and $\pm \phi_m$ are the positions of
the minima of the symmetric double well potential. The great
advantage of this formulation of the WKB approximation, beyond its
validity for ground state splittings, is that the limits of the
integrals are at the true extrema of the potential rather than the
classical turning points, making the calculation more tractable.

By evaluating Eqs. (\ref{eq:wkb_garg})-(\ref{eq:wkb_A})
numerically, we obtain a second result for $\Omega$, shown in Fig.
\ref{rf_SQUID_ground_state numerical}(a) as a function of
$\beta_L$. We see that the two forms of the WKB approximation are
similar in overall shape, with $\Omega$ vanishing at $\beta_L=1$
where $\omega_0$ becomes zero, and decreasing exponentially at
larger values of $\beta_L$.  However, the two forms disagree
quantitatively at small values of $\beta_L$ and diverge from one
another at large values of $\beta_L$.  These difficulties are
hardly surprising, since the WKB approximation assumes a
well-defined state localized in each well, and for states very
close to the top of the barrier this assumption is no longer
valid.  Thus, to obtain a more accurate tunneling frequency we need a
full quantum mechanical solution for the degenerate double-well
potential.

To find the wavefunctions we first choose a set of basis functions
${b_i(\phi)}$.  By calculating the Hamiltonian matrix elements
\begin{eqnarray}
H_{mn} &=& \int_{-\infty}^{\infty} b_n(\phi) H(\phi) b_m(\phi)
d\phi
\end{eqnarray}
and the overlap matrix
\begin{eqnarray}
B_{mn} &=& \int_{-\infty}^{\infty} b_n(\phi) b_m(\phi) d\phi,
\end{eqnarray}
we can find the energy levels as the eigenvalues of the matrix
\begin{eqnarray}
K &=& B^{-1} H.
\end{eqnarray}
To solve for the ground state wavefunction we choose as our basis
set $12$ simple harmonic oscillator wavefunctions centered in the
left well and $12$ more centered in the right well.  We use the
Hamiltonian
\begin{eqnarray}
H(\phi) &=& \frac{{\Phi_0}^2}{8\pi^2 L} \left[2 \beta_L
    \cos(\phi)+\left(\pi+\phi+\phi_x\right)^2 \right] \nonumber\\
    &&+ \frac{{\Phi_0}^2 C}{8\pi^2} \left(\frac{\partial}{\partial\phi}\right)^2,
\end{eqnarray}
where $\phi_x\equiv 2\pi\Phi_x/\Phi_0$.  The results for $\phi_x=0$
are shown in Fig. \ref{rf_SQUID_ground_state numerical}(a). For large
values of $\beta_L$ the full solution approaches the modified WKB
expression, Eq. (\ref{eq:wkb_garg}), asymptotically. As $\beta_L$ is
decreased toward unity the tunneling rate approaches a constant value.
This is in contrast to the semi-classical models which predict a
tunneling rate proportional to $\beta_L$ as the prefactor $\omega_0$
dominates; the full solution shows that this is an artifact of the
approximation.

Figure \ref{rf_SQUID_ground_state numerical}(b) shows $\Lambda$ {\it
  vs.} $\beta_L$ for the three calculations.  The two semi-classical
approximations predict that $\Lambda$ vanishes at certain values of
$\beta_L$, but this is an artifact of the apparent maxima in Fig.
\ref{rf_SQUID_ground_state numerical}(a); the full quantum treatment
shows no zero.  Figure \ref{rf_SQUID_ground_state numerical}(c) plots
the fractional change in tunneling frequency, $\delta\Omega/\Omega$,
{\it vs.} $\beta_L$ for the three calculations for three fractional
changes in critical current, $\delta I_0/I_0$.  We note that for
$\beta_L\gtrsim 1.1$ the three approaches differ by no more than a
factor of about two.

\begin{figure}
  \centering
    \caption{\it note: figure attached} {Three quantities for the ground state of the one-junction
    flux qubit at the degeneracy point calculated using the standard
    WKB approximation (solid), WKB approximation corrected for the
    ground state (dashed), and numerical solution for the
    wavefunctions (points), plotted as a function of the dimensionless
    screening parameter $\beta_L$. (a)
    Splitting frequency between ground and first excited states,
    (b) sensitivity parameter $\Lambda$,
    and (c) effects of critical current
    fluctuations of three magnitudes on tunneling rate.
    Parameters are from Friedman {\it et al.}: $L=240$ pH
    and $C=104$ fF. \cite{Lukens}
    \label{rf_SQUID_ground_state numerical}}
\end{figure}

\subsection {One-Junction Flux Qubit (Excited States)}

The first demonstration of a one-junction flux qubit did not employ
ground states, however, but excited states in deep, tilted potential
wells. \cite{Lukens} The WKB approximation is again unsuitable, for
two main reasons.  First, treating asymmetric potentials is more
difficult, because of different prefactors for the two wells, but this
can be overcome. \cite{Averin} More importantly, resonant tunneling,
which causes a dramatic increase in the tunneling rate when two energy
levels are aligned, is entirely absent from the WKB approximation.
Thus, the only way to calculate the sensitivity to critical current
fluctuations is to solve the Schr\"{o}dinger equation for the energy
levels numerically.

We adopt the approach of Sec. IV.A with a different basis set.  We use
$60$ harmonic oscillator wavefunctions centered between the minima of
the two wells, so that $B$ becomes the identity matrix.  To reproduce
the experimental conditions, \cite{Lukens} we set $\beta_L=1.5$ and
find the energy levels for successive values of applied flux $\phi_x$.
We find that the energy difference between the third and ninth excited
states has a local minimum at $\phi_x\approx 0.514\times2\pi$,
corresponding to the condition for resonant tunneling.  Fixing
$\phi_x$ at this value and sweeping $\beta_L$, we calculate the
relevant quantities for low frequency critical current fluctuations.
The results are shown in Fig. \ref{rf_SQUID_excited_state}.

In Fig. \ref{rf_SQUID_excited_state}(a) we see that near the resonant
point $\beta_L=1.5$, $\Omega$ decreases with increasing barrier
height, as one would expect from a semi-classical analysis, but
reaches a local minimum at a slightly higher value.  As $\beta_L$ is
increased further, $\Omega$ increases because the energy levels are no
longer resonant.  At the minimum, the derivative quantity $\Lambda$
vanishes, as the changing barrier height balances the loss of
resonance, indicating that the system is immune to small critical
current fluctuations at this point.  We note that on resonance, where
$\Lambda$ is almost optimally bad, the system is immune to flux noise,
because the energy is a minimum as a function of flux.  Thus, one can
exchange sensitivity to critical current fluctuations for sensitivity to
flux noise as appropriate.

\begin{figure}
\centering
    \caption{\it note: figure attached} {Numerical solution for the excited states of an asymmetric
    one-junction flux qubit.  (a) Tunneling
    frequency between the third excited state in the shallow well and
    the ninth excited state in the deep well as a function of
    $\beta_L$ for a system on resonance at $\beta_L=1.5$.  (b) Derived
    sensitivity to critical current
    fluctuations.  Device parameters are as in Fig.
    \ref{rf_SQUID_ground_state numerical}.
    \label{rf_SQUID_excited_state}}
\end{figure}

\subsection {Three-junction flux qubit}

The three-junction qubit consists of three Josephson junctions of
critical currents $I_0^a$,$I_0^b$, and $I_0^c$ in series in a
superconducting loop of geometric inductance $L$, as shown in Fig.
\ref{three junction SQUID}(a). \cite{Mooij-PRB,Mooij,Chiorescu} The
smallest of the junctions, $c$, primarily controls the barrier
height while the larger two junctions, $a$ and $b$, serve as Josephson
inductors.  We parameterize this device by the ratios of the
Josephson coupling energy of the three junctions to the charging
energy $E_C=e^2/2C$, where $C$ is the mean capacitance of the two
larger junctions: $E_J^{a,b,c}/E_C=I_0^{a,b,c}\Phi_0/2\pi
E_C=\gamma^{a,b,c}$.  We assume that the junctions are in the
phase regime where $\gamma^{a,b,c}>>1$ and require that
$1/2<2\gamma^c/(\gamma^a+\gamma^b)<1$ so that a double-well
potential is formed. We consider the junctions individually so
that we may allow their critical currents to fluctuate
independently, and consider the case where asymmetries in the
large junctions are small, {\it i.e.}
$2\gamma^b/(\gamma^a+\gamma^b)<<1$. The energy landscape at
applied flux $\Phi_0/2$ exhibits multiple wells, most notably two
degenerate wells separated by a tunnel barrier that is much lower
than the barriers to all other flux states. The potential can be
written
\begin{equation}
  V(\delta)=(E_C/8C)(\gamma^a+\gamma^b+4\gamma^c \cos \delta)^2,
\end{equation}
where $\delta$ is a variable aligned with the tunneling direction
that is derived from the three junction phases.  In the
small-inductance limit, we can apply the WKB approximation given
in Eqs. (\ref{eq:wkb_garg})-(\ref{eq:wkb_A}) to calculate the rate
for this so-called intracell tunneling
\begin{widetext}
\begin{eqnarray}
\Omega&=& \frac{\Gamma E_C}{\hbar}
\exp\left[-\frac{(4\gamma^c+\gamma^a+\gamma^b)\left\{\sqrt{
        (4\gamma^c)^2-(\gamma^a+\gamma^b)^2}-(\gamma^a+\gamma^b)
        \arccos \left( \frac{\gamma^a+\gamma^b}{4\gamma^c} \right) \right\}}{2 \sqrt{\gamma^c(\gamma^a+\gamma^b)
        (4\gamma^c+\gamma^a+\gamma^b)}}\right], \label{eq:omega-Ej}
\end{eqnarray}
where
\begin{eqnarray}
\Gamma=\frac{(4\gamma^c-\gamma^a-\gamma^b)^{5/4}(\gamma^a+\gamma^b)^{1/4}(4\gamma^c+\gamma^a+\gamma^b)}{2\pi^{1/2}(\gamma^c)^{7/4}}.
\end{eqnarray}
\end{widetext} We note that the exponent reduces to a form
previously obtained \cite{Mooij-PRB} when $\gamma^a=\gamma^b$;
however the prefactor differs.

\begin{figure}
\centering
 \caption{{\it note: figure attached} Three-junction flux qubit.  (a) Schematic showing inductive
   loop, embracing $\Phi_0/2$ interrupted by three Josephson
   junctions.  (b) Tunneling frequency and (c) $\Lambda$ {\it vs.}
   Josephson-to-charging energy ratio.  Solid lines indicate
   dependence on large junction ratio $\gamma^{a,b}$ with
   $\gamma^c=28$, and dashed lines indicate dependence on small
   junction ratio $\gamma^c$ with $\gamma^a=\gamma^b=35$.  $E_C=7.4$
   GHz for all plots.
\label{three junction SQUID}}
\end{figure}

To calculate the effects of low frequency noise, we must
account for the fact that the critical currents of the three junctions
fluctuate independently.  Because the small and large junctions
play different roles, we consider changes in each separately.  We
adopt parameters used in the experiments of Chiorescu \textit{et
al.}, \cite{Chiorescu} $\gamma^a=\gamma^b=35$, $\gamma^c=0.8\times
\gamma^{a,b}=28$, and $E_C/2\pi\hbar=7.4$ GHz.  In Fig. \ref{three
junction SQUID}(b), we plot the tunneling frequency $\Omega /2\pi$
as a function of the Josephson-to-charging energy ratios for each
of the three junctions holding the other two constant.  Figure
\ref{three junction SQUID}(c) shows $\Lambda_i=(\gamma^i/\Omega)
\partial\Omega/\partial\gamma^i$, where
%$i\in\{a,b,c\}$,
$i=a, b$ or $c$, as a
function of the same variables. For the experimental parameters,
we calculate $\Omega/2\pi=7.96$ GHz, which differs somewhat from
the experimentally observed value of $3.4$ GHz; however the
exponential dependence in Eq. (\ref{eq:omega-Ej}) magnifies
parametric uncertainties, making exact agreement unlikely.  We see
that the small junction is indeed the dominant contribution to
$\Lambda$, with $\Lambda_{a,b}=4.6$ and $\Lambda_c=10.4$.  Adding
the contributions incoherently gives
$\Lambda=(\Lambda_{a}^2+\Lambda_b^2+\Lambda_{c}^2)^{1/2}=12.3.$

\subsection {Single Josephson junction (phase) qubit}

Martinis and coworkers have used a single, current-biased Josephson
junction as a qubit, the $|0\rangle$ and $|1\rangle$ states being the
ground and first excited states of the tilted washboard potential
well, as shown in Fig. \ref{single junction}(a). The energy
separation between energies $E_0$ and $E_1$ is
\begin{eqnarray}
\Omega &=& (E_1-E_0)/\hbar \approx
\omega_p,\label{eq:omega-diff}
\end{eqnarray}
where
\begin{eqnarray}
\omega_p &=& \left(2 \sqrt{2} \pi I_0/C \Phi_0\right)^{1/2}
\left( 1-I/I_0\right)^{1/4}\label{eq:omega-p-jj}
\end{eqnarray}
is the small oscillation (plasma) frequency in the well.  In Fig.
\ref{single junction}(b) we plot $\Omega$ {\it vs.} $I/I_0$ for the
parameters used in the experiments of Martinis \textit{et al.}
\cite{Martinis-Rabi}  We determine $\Lambda$ {\it vs.} $I/I_0$ from
Eq. (\ref{eq:omega-p-jj}), and plot the result in Fig. \ref{single
  junction}(c).  At the bias point used in the experiments, $I=20.77$ $\mu$A ($I/I_0=0.985$), $\Lambda$ has the value $16$ at a
tunneling frequency $\Omega/2\pi = 6.9$ GHz.

\begin{figure}
\centering
 \caption{{\it note: figure attached} Single Josephson junction qubit.   (a) Schematic and (b) energy
   level diagram. (c) Variation of energy separation with bias
   current. (d) $\Lambda$ as a function of bias current. Parameters
   are from Martinis {\it et al.}: $C=6$ pF,
   corresponding to a junction area of $100$ $\mu$m$^2$, and
   $I_0=21.1$ $\mu$A. \cite{Martinis-Rabi} \label{single junction}}
\end{figure}

\subsection {Quantronium (hybrid charge-flux) qubit}

The qubit developed by the Saclay group consists of a Cooper pair box,
a small island coupled by Josephson junctions of critical current
$I_0$ and capacitance $C_j$ on each side, connected in a
superconducting loop containing a Josephson junction with a much
larger critical current
[Fig. \ref{quantronium}(a)]. \cite{Quantronium}
The island is connected to a voltage source {\it
  via} a capacitor $C_g$.  The circuit parameters are selected with
the Josephson energy $E_J^{a,b}=\Phi_0 I_0^{a,b} / 2 \pi$ comparable
to the charging energy $E_{CP}=(2e)^2/2 (C_g+2 C_j)$, so that the
device operates in the crossover regime between the charge and flux
modes.  In this configuration, a charge induced on the central island
generates a phase change around the loop, driving a circulating current
determined by the Josephson inductance of the two small junctions.
This current is detected by measuring the pulsed current required to
exceed the critical current of the readout junction, $I_0^r$.  The
qubit energy levels $E_0$ and $E_1$ are controlled by the gate charge
$N_g e$ and the phase difference $\delta$ across both junctions
according to the approximation \cite{Cottet}
\begin{eqnarray}
E_{0,1} &=& \mp \left\{ \left[ \frac{E_J}{2} \cos
\left(\frac{\delta}{2}\right)\right]^{2} + \left[ E_{CP} (1-2N_g)
\right]^{2} \right\} ^{1/2}.\label{eq:quantronium-E}
\end{eqnarray}
where $E_J=E_J^a+E_J^b$ is the total Josephson coupling energy.
Thus, the qubit frequency, which is proportional to the level
spacing, is
\begin{eqnarray}
\hbar \Omega &=& E_1-E_0 \\
&=& 2\left\{ \left[ \frac{E_J}{2} \cos
\left(\frac{\delta}{2}\right)\right]^{2}+ \left[ E_{CP} (1-2N_g)
\right]^{2} \right\} ^{1/2}.\label{eq:quantronium-freq}
\end{eqnarray}
When $N_g$ and $\delta$ are adjusted to the optimal working point,
$\delta = 0$ and $N_g = 1/2$, the system is maximally insensitive to
phase and charge fluctuations; however, incoherent fluctuations in the
critical current of the small junctions couple linearly to the level
splitting without perturbing the phase or charge to first order,
giving $\Lambda=2^{-1/2}$.  Away from $N_g=1/2$, $\Lambda$ is
reduced, as plotted in Fig. \ref{quantronium}(b) for the parameters
used in the Saclay experiments, $C_j=2.7$ fF ($E_{CP}/k_B=0.68$ K), and
$I_0=18$ nA [$(E_J^a+E_J^b)/k_B=0.86$ K], but the device is then no longer immune
to charge fluctuations.

\begin{figure}
  \centering  \caption{{\it note: figure attached} The quantronium qubit, which operates in the crossover regime
   between the charge and flux modes, converts charge oscillations
   on the single electron transistor to flux modulation in the loop.
   (a) Schematic showing phase difference $\delta$ across two small
   Josephson junctions with charge $N_g$ on island between them.  (b)
   Level splitting frequency $\Omega /2\pi$ and (c) critical current
   sensitivity $\Lambda$ {\it vs.} $N_g$.  Curves are plotted for the
   parameters reported by Vion {\it et al.}, $I_0=18$ nA, $C_j=2.7$ fF;
   at the optimal working point $N_g =1/2$, $\delta=0$,
   $\Lambda=2^{-1/2}$, and $\Omega$ is calculated to be $17.9$ GHz,
   slightly different from the observed value of $16.5$ GHz.}
 \label{quantronium}
\end{figure}

\section {1/f CRITICAL CURRENT FLUCTUATIONS}

Critical current fluctuations in Josephson junctions have been
extensively studied over the past two decades, mostly to understand
the low frequency noise in SQUIDs.  As a result, most of the reported
measurements have been in the temperature range $1-4$ K on junctions of
areas from $4-100$ $\mu$m$^2$.  We first briefly describe scaling of the
data by the junction area, the critical current, and temperature.

As mentioned earlier, it is generally accepted that critical
current noise in Josephson junctions arises from charge trapping
at defect sites in the barrier.  A trapped charge locally modifies
the height of the tunnel barrier, changing the resistance of the
junction, and, in the case of a Josephson junction, also the
critical current.  For a junction of area A, the change in
critical current is $\Delta{I_0} = (\Delta A/A)I_0$, where $\Delta
A$ is the effective area of the junction over which tunneling is
blocked by the temporary presence of the trapped charge.  The
critical current spectral density for one trap is proportional to
$(\Delta I_0)^2$, so that the spectral density for $N$ identical,
independent traps scales as $N(\Delta I_0)^2 = nA(\Delta A/A)^2
{I_0}^2$, where $n$ is the number of traps per unit area.
Consequently, for a given junction technology characterized by a
trap density $n$ and blocking area $\delta A$, we expect the
critical current spectral density $S_{I_0}(f)$ to scale as
$I_0^2/A$.  To test this hypothesis, we have compiled a series of
measurements of the $1/f$ critical current noise at temperature
$T=4.2$ K, taken in a variety of junctions and dc SQUIDs by
different groups (Table I).  For each, we list the critical
current $I_0$ and area $A$ of the junctions, which vary by several
orders of magnitude, and the magnitude of the critical current
noise spectral density at $1$ Hz, $S_{I_0}(1$ Hz$)$.  We observe
that the quantity $S_{I_0}^{1/2}(1$ Hz$)A^{1/2}/I_0$ is remarkably
constant, varying by less than a factor of $3$.

This result supports the charge trap model for the $1/f$ critical
current noise, and, since it includes measurements on different
junction barrier materials (AlOx, InOx, NbOx)
even suggests that the product of the trap density and Coulomb
screening area must be similar in magnitude for these different
oxides.

Averaging these measurements, we estimate the critical current noise
at $4.2$ K for any junction of critical current $I_0$ and area $A$ to
be
\begin{eqnarray}
S_{I_0} \left( 1 {\rm Hz},4.2 {\rm K} \right) &\approx& 144 \frac{\left( I_0/
\mu\,{\rm A} \right)^2}{A/ \mu {\rm m}^2}
\frac{({\rm pA})^2}{{\rm Hz}}.\label{eq:S-I0}
\end{eqnarray}
The temperature dependence of the $1/f$ critical current noise is
less firmly established.  Since the charge traps responsible for
the noise are thought to be in the tunneling regime at low
temperatures, one might expect that the temperature dependence
would be weak.  However, the only measurement of the spectral
density of the critical current noise in Josephson junctions at
low temperatures we are aware of showed a $T^2$ dependence from
$4.2$ K down to about $300$ mK. \cite{Wellstood-thesis} The issue
of whether or not this behavior extends to lower temperatures is
of crucial importance to the development of qubits involving
Josephson junctions.

In the absence of other data or models, we take the optimistic
view that $S_{I_0}(f,T)$ scales quadratically with temperature and so
is dramatically reduced at the low temperatures where
superconducting qubits are operated.  We thus take as a working
hypothesis
\begin{eqnarray}
S_{I_0} \left( f,T \right) &\approx& \left[144
\frac{\left( I_0/ \mu\,{\rm A} \right)^2}{\left( A/ \mu\, {\rm
m}^2 \right)} \left(
  \frac{T}{4.2 {\rm K}}
\right)^2 ({\rm pA})^2 \right] \frac{1}{f}.~~~~\label{eq:S-I0-f}
\end{eqnarray}

The observed $T^2$ dependence is incompatible with the electron
trapping mechanism in the tunneling regime, which predicts a linear
temperature dependence. \cite{Dutta-Horn} There is strong evidence
that charge trapping occurs via tunneling in the temperature range
considered, so that the noise should be relatively temperature
independent.  Further, for $eV$, $k_BT << 2 \Delta$, where $\Delta$ is
the energy gap, both the available number of single electrons and the
available number of final single-electron states scale as $\exp(-\Delta
/k_BT)$, so that charge trapping is expected to freeze out at low
temperatures.  This leads one to seek alternative explanations.  One
possibility is that the $1/f$ noise is associated with leakage
currents at voltages below $2 \Delta / e$, which do not exhibit an
exponential temperature dependence.  Such leakage currents presumably
occur between opposing normal regions of the electrodes, conceivably
at the edges of the junctions or along the core of a flux vortex
penetrating the junction.  An investigation of the correlation between
leakage currents and $1/f$ noise would be of great interest.  Other
possible sources of the $1/f$ noise include the motion of electrons
between traps within the tunnel barrier, and the motion of vortices in
or near the junction, which could create a thermally-activated
contribution to the critical current fluctuations.  We note that a
thermally activated model yielding a $T^2$ dependence has been
proposed by Kenyon {\it et al.} \cite{Kenyon} in the context of charge
$1/f$ noise, but should be equally applicable to critical current
noise.  In this model, one assumes that the two-state systems have
asymmetric wells, and that the depths of the wells are independent
random variables.

\par
\begin{table*}[] \caption{Compilation of $1/f$ critical
current noise measurements in Josephson junctions of different
technologies, areas $A$, and critical currents $I_0$ at $4.2$ K;
$S_{I_0}(1$ Hz$)$ is the spectral density at $1$ Hz.  The relative
invariance of the scaled quantity {$A^{1/2}S_{I_0}^{1/2}(1$
Hz$)/I_0$} supports the charge trapping mechanism for the $1/f$
noise.}\label{1/f noise data table} \addvspace{6pt}

\begin{tabular}{ccccc}
\hline
Junction&$\hspace{20pt}A\hspace{20pt}$&$\hspace{20pt}I_0\hspace{20pt}$&$\hspace{20pt}S_{I_0}^{1/2}(1$ Hz$)\hspace{20pt}$&$A^{1/2}S_{I_0}^{1/2}(1$ Hz$)/I_0$\\
technology&$\mu$m$^2$&$\mu A$&pA$/$Hz$^{1/2}$&$\mu$m$($pA$/$Hz$^{1/2})/\mu$A\\
\hline\hline
Nb-AlOx-Nb\cite{Savo}&9&9.6&36&11\\
 &8&2.6&6&7\\
 &115&48&35&8\\
 &34&12&41&20\\
\hline
Nb-NbOx-PbIn\cite{Wellstood-thesis}&4&21&74&7\\
 &4&4.6&46&20\\
 &4&5.5&25&9\\
 &4&5.7&34&12\\
 &4&11.4&91&16\\
\hline
Nb-NbOx-PbInAu\cite{Foglietti}&1.8&30&184&8\\
\hline
PbIn-InOx-Pb\cite{Koch-Pb}&6&510&3300&15\\
\hline
\hline
Average& & & &12\\
\hline
\end{tabular}
\end{table*}
\par

\section {DETERMINATION OF DEPHASING TIMES}

As described above, the low frequency critical current
fluctuations generate phase noise and decoherence in any
measurement of quantum coherent oscillations. To determine the
effect of the fluctuations on $\tau_{\phi}$, we simulate the
oscillations of the qubit state probability distribution.

In general, there are two techniques for observing quantum
oscillations in superconducting qubits. The qubit bias can be pulsed
suddenly to the degeneracy point where the qubit oscillates between
the measurement basis states at frequency $\Omega$. After time $t$,
the qubit bias is pulsed suddenly away from the degeneracy point,
after which the measurement is performed. \cite{Nakamura1}  In this
section we consider such a degeneracy point measurement for a
superconducting qubit in the presence of low frequency critical
current fluctuations. We normalize the qubit states to $+1$ and $-1$
and always initialize the state to $+1$ before each bias pulse to the
degeneracy point.  For qubits coupled to Ohmic dissipation and without
critical current fluctuations, the subsequent oscillations of the
expectation value $\langle Z(t) \rangle$ decay with the dephasing time
$\tau_{\phi}^0$ according to
\begin{eqnarray}
\langle Z(t) \rangle &=& e^{-t/\tau_{\phi}^0}\cos \Omega t.\label{eq:prob-not}
\end{eqnarray}
We will see that the low frequency noise provides an additional
mechanism for decoherence and a different functional form for the
decay of $\langle Z(t) \rangle$.

Alternatively the qubit bias can remain fixed away from the
degeneracy point while the qubit is driven between the ground and
excited states with resonant microwave pulses of varying width.
This technique has been used to measure Rabi oscillations of the
quantum state in several superconducting qubits.
\cite{Quantronium,Chiorescu,Martinis-Rabi}  A measurement of the
dephasing time $\tau_{\phi}$ in this driven case requires a more
sophisticated pulse arrangement, such as a Ramsey fringe sequence.
\cite{Quantronium, Chiorescu}  We note that for the single
Josephson junction phase qubit, \cite{Martinis-Rabi}  resonant
microwave driving is the only possible technique for observing
quantum oscillations as there is no degeneracy point at which the
qubit can be operated. Nonetheless, we expect our calculation of
the dephasing due to critical current fluctuations from a
simulation of an experiment involving switching to and away from
the degeneracy point to give a reasonable estimate for
$\tau_{\phi}$ in the microwave-driven experiments as well.

For our simulations of the quantum oscillations at the degeneracy
point, we allow the qubit to evolve for time $t$ followed by a
single-shot measurement with a sampling window that is much
shorter than $2 \pi / \Omega$ (Fig. \ref{sampling_methods}). We
assume that the interval between consecutive single-shot
measurements of the state is $t_{Z}$; this interval includes the
time to initialize the state, the delay time during which the
qubit evolves, the sampling time, the readout time, and any time
allotted for the system to thermalize following the dissipative
measurement.  To map out the time dependence of the qubit state,
we measure the expectation value $N_t$ times, at intervals
separated by time $t_d$, each point being the average of $N_{Z}$
measurements. From this time evolution, we can determine the
envelope and its characteristic decay time, and, if
the sampling frequency is above the Nyquist frequency (twice the
coherent oscillation frequency), the oscillation frequency. The
key point is that low frequency fluctuations in the critical
current cause the oscillation frequency to be different for each
successive single-shot measurement of the qubit, resulting in an
effective dephasing.

Because of the nature of $1/f$ noise, the resulting dephasing
depends both on the total number of samples $N = N_Z N_t$ (which
sets the elapsed time of the experiment $N t_Z$) and on the
sequence in which the measurements are taken. We consider two
cases, illustrated in Fig. \ref{sampling_methods}. Method A is
time-delay averaging, in which we take $N_Z$ successive
measurements for each time delay and average them to find the
qubit expectation value at that delay time. Method B is time-sweep
averaging, in which we make a single measurement at each of the
$N_t$ points, and then average $N_Z$ such time sweeps to generate
the qubit time evolution. These differ because of the time scales
involved in $1/f$ noise: Method A averages only high frequency
fluctuations at each time-delay point, while Method B averages
both high and low frequency components. Data sampling schemes
intermediate between these extremes are also possible; these
involve the averaging of $N_s < N_Z$ multiple sweeps, each
acquired by sampling $N_m = N_Z/N_s$ successive measurements at
each time delay value.

\begin{figure}[b]
\centering
  \caption{{\it note: figure attached} Measurement sequences for mapping out coherent
oscillations.  (a) Method A: time-delay averaging.  (b) Method B:
time-sweep averaging.  The interval between qubit state
measurements is $t_Z$; the spacing of time-delay points is $t_d$.
\label{sampling_methods}}
\end{figure}

For method A, the expectation value after time $t_m = m t_d$, with
$1\leq m\leq N_t$, is given by
\begin{eqnarray}
\langle Z^A (t_m) \rangle  & = &  \frac{1}{N_Z}  \sum_{n=1}^{N_Z}
\cos \left\{ \left[ \Omega + \frac{d \Omega}{dI_0} \delta I_0
\left(  t_A \right) \right]
t_m \right\}  e^ {-\frac{t_m}{\tau_{\phi}^0}} \nonumber \\
&=& \frac{1}{N_Z}  \sum_{n=1}^{N_Z}
 \cos \left\{  \Omega \left[ 1+ \Lambda \delta i_0
\left( t_A \right) \right] t_m \right\} e^
{-\frac{t_m}{\tau_{\phi}^0}}, \nonumber \\
&& \label{eq:prob-aveA}
\end{eqnarray}
where $t_A=\left[ \left( m-1 \right) N_Z + n \right] t_Z$.
For method B we have
\begin{eqnarray}
\langle Z^B (t_m) \rangle  & = &  \frac{1}{N_Z}  \sum_{n=1}^{N_Z}
 \cos \left\{ \left[ \Omega + \frac{d \Omega}{dI_0} \delta I_0
\left( t_B \right) \right] t_m \right\}  e^
{-\frac{t_m}{\tau_{\phi}^0}} \nonumber \\
& = &  \frac{1}{N_Z}  \sum_{n=1}^{N_Z}
 \cos \left\{  \Omega \left[ 1+ \Lambda \delta i_0
\left( t_B  \right) \right] t_m \right\}  e^
{-\frac{t_m}{\tau_{\phi}^0}}, \nonumber \\
&& \label{eq:prob-aveB}
\end{eqnarray}
where $t_B=\left[ \left( n-1 \right) N_t + m \right] t_Z$. Here
$\tau_{\phi}^0$ is the dephasing time set by decoherence
mechanisms besides $1/f$ noise such as dissipative processes in
the qubit and the environment. To simulate the dephasing due to
critical current fluctuations alone, we take $\tau_{\phi}^0$ to be
infinite. The quantity $\delta{I_0(t)}$ is the time-varying
deviation in the critical current from its average value. Note
that the changes in oscillation frequency scale with $\Lambda$ and
with the fractional changes in the critical current $\delta
i_0(t)=\delta{I_0(t)}/I_0$.

We determine the time sequence of critical current fluctuations
(Fig. \ref{decoherence_simulation}) by Fourier transforming a
complex spectrum of critical current fluctuations. This spectrum
is generated in frequency space, with magnitudes randomly chosen
from an exponential distribution with a mean value equal to
$(S_{I_0}(1$ Hz$)/f)^{1/2}$ and phases randomly chosen from a
uniform distribution from $0$ to $2\pi$. This procedure is
equivalent to sampling real and imaginary components of the
critical current fluctuations from Gaussian distributions centered
at zero magnitude, thus ensuring that the generated noise is
Gaussian. The actual critical current fluctuations of the junction
may not be strictly Gaussian if interactions between the charged
traps are present, but the assumption of Gaussian statistics
should give a good representation of the noise. The relevant
frequency range is from $f_{max}=1/t_Z$, set by the single-shot
measurement time, to $f_{min}=1/N t_Z$, where $N t_Z$ is the total
duration of the experiment.  As an example, consider an experiment
in which $t_Z=1ms$, $N_Z=100$, and $N_t=100$.  We generate $N =
10^4$ time sequence points over the period $N t_z=10$ s.  We
choose a representative qubit with a junction of critical current
$I_0=1$ $\mu$A and area $A=0.01$ $\mu$m$^2$.  At $T=100$ mK, the
universal $1/f$ noise spectral density from Eq.(\ref{eq:S-I0-f})
yields $S_{I_0}(1$ Hz$)=8.16\times10^{-24}$A$^2$Hz$^{-1}$,
corresponding to a root-mean-square fractional change in the
critical current of about $10^{-5}$ over the bandwidth from
$10^{-1}$ to $10^3$ Hz. Figure \ref{decoherence_simulation}(a)
shows a typical time trace simulated with these parameters.  The
enhanced low frequency components present in the $1/f$ spectrum
are evident in the fluctuation spectrum.

\begin{figure}
\centering
\caption{({\it note: figure attached} a) Simulated time-sequence of critical current changes
for an experiment with $N=10^4$ total qubit state measurements
taken at intervals of $t_Z=1$ ms. (b) Corresponding $1/f$ frequency
spectrum. \label{decoherence_simulation}}
\end{figure}

To simulate the observed coherent oscillations, we insert such a
noise time-sequence of the required duration into
Eqs.(\ref{eq:prob-aveA}) and (\ref{eq:prob-aveB}).  In Fig.
\ref{probability_envelopes}, we show the probability amplitude
${\langle Z \rangle}$ calculated for $N_t=1000$ time delay points,
each averaged over $N_Z=3000$ qubit state measurements (thus, $N=3
\times 10^6$) acquired by sampling methods A and B.  We assume the
qubit parameters $I_0=1$ $\mu$A, $A=0.01$ $\mu$m$^2$, $\Omega
/2\pi=1$ GHz and $\Lambda=100$, with $T=100$ mK.  The optimum
sampling rate is larger than the Nyquist frequency so that the
characteristic qubit oscillation frequency can be determined, and
incommensurate with the oscillation period of the qubit, so that
the envelope of the oscillations is fully delineated and not
aliased.  In this case, we arbitrarily choose the sampling
frequency to be the irrational number $(1+\phi)\Omega/2\pi \approx
2.618$ GHz, where $\phi=(1+\sqrt 5)/2 \approx 1.618$ is the Golden
mean.  Thus, $t_Z=0.382$ ns. The envelope function is calculated
by demodulating the oscillations {\it via} convolution of the
averaged probability amplitudes with the Gaussian filter kernel
\begin{eqnarray}
K(t) & = & \left( \frac{1}{2 \pi \sigma^2}\right)^{1/2} \exp
\left( -t^2 / 2 \sigma^2 \right) ,\label{kernel}
\end{eqnarray}
where $\sigma$ is chosen to be the sampling period $t_Z$.

The oscillation amplitude of the qubit state is found to decay
with a Gaussian envelope function
\begin{eqnarray}
{\langle Z \rangle}_{env} &\sim& \exp \left( -t^2 / 2
\tau_{\phi}^2 \right) ,\label{eq:phi-env}
\end{eqnarray}
where $\tau_{\phi}$ is a characteristic dephasing time.  This form
arises from the frequency modulation of the qubit by the critical
current fluctuations, in contrast to an exponential decay induced
by dissipative processes.  We note that for long delay times the
envelope does not vanish but instead saturates to a noise floor
level that corresponds to uniform randomization of the oscillation
phase by the critical current fluctuations.  The noise floor is
$Z_{noise} \sim N_Z^{-1/2}$ for both Methods A and B. Particularly
for small $N_Z$, it is necessary to account for the noise floor to
make an accurate determination of $\tau_\phi$. We do this by
fitting to
\begin{eqnarray}
{\langle Z \rangle}_{env} &\sim& Z_{noise}+(1-Z_{noise}) \exp
\left( -t^2 / 2 \tau_{\phi}^2 \right).~~~~
\label{eq:phi-env-noise}
\end{eqnarray}
Both the dephasing times and the scatter in the amplitude envelope
are different for the two methods.  Method A gives a longer
dephasing time than Method B, in this case by about $30\%$. This
occurs because all of the qubit state measurements at a particular
delay time for Method A are acquired in a time interval $N_Z t_Z$,
rather than over the entire experiment duration $N t_Z$ as in
Method B. Thus, the number of decades of $1/f$ noise that affect
the qubit dynamics in Method A is $\log(N_Z)=3$, compared to
Method B which samples $\log(N)=6$ decades. The scatter in the
simulated data is also greater for Method A because the low
frequency variation of the tunneling frequency is not averaged
out.  The origin of this scatter can be best understood by
choosing junction and measurement parameters for which $\tau_\phi$
and $T_{osc}$ are comparable so that the coherent oscillations and
the amplitude decay can be resolved simultaneously. In Fig.
\ref{probability_oscillations}, we show the probability amplitude
for the same qubit parameters but with a substantially increased
level of critical current fluctuations, approximately $40$ times
larger in amplitude, calculated for $N_t=200$. Here, the discrete
oscillations are clear for Method B but quite distorted for Method
A.  The dephasing time for Method A is again longer, in this case
by about $22\%$.

\begin{figure}
\centering
  \caption{{\it note: figure attached} Probability envelopes determined by simulations using
  measurement Methods A and B for a qubit with $I_0=1$ $\mu$A,
  $S_{I_0}(1$ Hz$)=8.16\times 10^{-24}$ A$^2$Hz$^{-1}$,
  $A=0.01$ $\mu$m$^2$, $\Lambda=100$, and $\Omega/2\pi=1$ GHz.
  The structure visible in the Method B plot arises from periodic
  sampling of the oscillations and is evidence of the
  increased effective averaging relative to Method A.
\label{probability_envelopes}}
\end{figure}

\begin{figure}
\centering
\caption{{\it note: figure attached} Simulated probability oscillations with large critical current
  fluctuations for measurement Methods A and B. Qubit parameters as in
  Fig. \ref{probability_envelopes}, except $S_{I_0}(1$ Hz$)=1.39\times
  10^{-20}$ A$^2$Hz$^{-1}$
  \label{probability_oscillations}}
\end{figure}

Because of the low frequency divergence of $1/f$ noise, the
variance in the measured dephasing time is substantial, and it is
necessary to carry out a series of experimental runs to determine
the dephasing time accurately for a given set of junction and
measurement parameters. The spread in dephasing times can be seen
in Fig. \ref{histogram} in which we plot distributions of the
dephasing times obtained by Methods A and B for the qubit
parameters used in Fig. \ref{probability_envelopes} and for
different numbers of flux measurements. For any value of $N$, the
mean dephasing time is larger for Method A than for Method B, as
expected since fewer decades of 1/f noise affect the qubit; the
standard deviations are larger for Method B.

With a series of such simulations for different junction and qubit
parameters, it is straightforward to establish that $\tau_{\phi}$
is proportional to $I_0$ and inversely proportional to $\Omega$,
$\Lambda$, and $S_{I_0}^{1/2}(1$ Hz$)$. The dependence of
$\tau_{\phi}$ on the number of measurements, which sets the range
of $1/f$ noise that is effective in dephasing the qubit, can be
found by carrying out the simulations for different measurement
parameters $N_t$ and $N_Z$, as shown in Fig. \ref{histogram}.  The
mean dephasing times for a series of simulations with the same
parameters described above are shown in Fig. \ref{N_dependence}.
As discussed above, Method A gives longer times than Method B for
all values of $N$. We find that the dephasing time $\tau_\phi$ for
Method A decreases as a weak power-law of $N$, which is expected
since the frequency range of the $1/f$ noise increases for larger
$N_Z$. For Method B, $\tau_\phi$ is nearly constant, changing by
only a few percent over 3 orders of magnitude in $N$. This
insensitivity likely arises because the increased frequency range
of the noise for larger $N$ (which should suppress the the
dephasing time) is compensated by the increased averaging which
smoothes the fluctuations.  For large $N$, $\tau_\phi$ for Method
B agrees well with the analytical result obtained by Martinis,
{\it et al.}, \cite{Martinis-noise} differing only by a numerical
factor of order unity, but deviates substantially at lower $N$.

\begin{figure}
\centering
  \caption{{\it note: figure attached} Distributions of dephasing times $\tau_\phi$ calculated
by Method A (open symbols) and Method B (closed symbols) for
different number of flux measurement points $N = 3 \times 10^4$
(squares), $3 \times 10^5$ (triangles), and $3 \times 10^6$
(circles). Each distribution includes $1000$ simulations of the
coherent oscillations accumulated into bins of width $2$ ns. Qubit
parameters are as in Fig. \ref{probability_envelopes}.
  \label{histogram}}
\end{figure}

\begin{figure}[b]
\centering
    \caption{\it note: figure attached} {Variation of the dephasing
time $\tau_\phi$ with the number of qubit state measurements $N$
for Methods A and B. Each point corresponds to the mean value of
$\tau_\phi$ from $50$ simulations of the oscillation decay
envelope. Qubit and noise parameters as in Fig.
\ref{probability_envelopes}. \label{N_dependence}}
\end{figure}

Using our empirical expression for $S_{I_0}(f)$, Eq.
(\ref{eq:S-I0-f}), and taking the number of qubit measurements in
a typical experiment to be $N=10^6$, we find
\begin{eqnarray}
\tau_{\phi}^A \left( \mu {\rm s} \right) &\approx& 20
A^{1/2}(\mu{\rm m})/\Lambda (\Omega/2 \pi)({\rm GHz}) T({\rm
K})~~~~
\end{eqnarray}
for sampling by Method A and
\begin{eqnarray}
\tau_{\phi}^B \left( \mu {\rm s} \right) &\approx& 15
A^{1/2}(\mu{\rm m})/\Lambda (\Omega/ 2 \pi)({\rm GHz}) T({\rm
K})~~~~
\end{eqnarray}
for Method B.

>From these results, we estimate the values of $\tau_{\phi}$ and
$\Omega \tau_{\phi}/2 \pi$ predicted for each of the qubit schemes
described in Sec. III, using the device parameters reported in the
experiments and assuming sampling by Method B with $N=10^6$. We have set $T=100$
mK and assumed explicitly that the $T^2$ dependence of
$S_{I_0}(f)$ extends to this temperature. These results are listed
in Table II. For comparison, we also list the measured dephasing
times and the temperatures at which the experiments were
performed.  Our estimated dephasing times range between $0.8$ $\mu$s and $12$ $\mu$s, with the longer times corresponding to the
qubit schemes with larger area junctions. Such times would allow
for several thousand oscillations of the quantum state, making
possible various quantum computing operations. However, with the
exception of quantronium, the measured dephasing times are orders
of magnitude shorter than our estimated values, indicating that
other sources of decoherence
are dominant. In the quantronium experiments, the isolation
obtained by operating at the optimal working point, described in
Section III.E, enhances the coherence time nearly to the value
where our estimates (at $100$ mK) predict critical current fluctuations would
have a noticeable effect; however, $S_{I_0}$ may be substantially
smaller at the experimental temperature of $15$ mK.

\par
%\begin{table*}[t]
\begin{table*}[t]
\caption{Estimated dephasing times at $100$ mK due to $1/f$ noise in
  $I_0$ for various qubit schemes. Measured dephasing times and
  experimental temperatures are
  included where measurements exist. For the one-junction flux qubit columns,
  values of $\Omega/2 \pi$ were calculated as described in the text. All
  other values of $\Omega/2 \pi$ were taken from corresponding
  experiments. Values of $\Lambda$ for each qubit scheme were calculated
  as described in Sec. III.} \label{qubit parameters}
%\begin{tabular}[h]{|c|c|c|c|c|c|}
\addvspace{6pt}
\begin{tabular}{cccccc}
\hline
\hspace{20pt}Parameter \hspace{20pt}&1-junction&1-junction&3-junction&single&\hspace{12pt}quantronium\cite{Quantronium}\hspace{12pt}\\
%\hspace{20pt}\hspace{20pt}&(ground state)&(excited state)\cite{Lukens}&SQUID\cite{Chiorescu}&&\hspace{12pt}\hspace{12pt}\\
\hspace{20pt}\hspace{20pt}&flux qubit&flux qubit&flux qubit\cite{Chiorescu}&junction\cite{Martinis-Rabi}&\hspace{12pt}\hspace{12pt}\\
\hspace{20pt}\hspace{20pt}&(ground state)&(excited state)\cite{Lukens}&\hspace{12pt}\hspace{12pt}\\
\hline\hline
$I_0(\mu$A$)$&1.46&1.46&0.5&21.1&0.018\\
$A(\mu$m$^2)$&2.0&2.0&0.05&100&0.02\\
$\Lambda$&40.6&71.5&12.3&16&0.7\\
$\Omega/2\pi($GHz$)$&3.4&0.59&3.4&6.9&16.5\\
calc $\tau_\phi(\mu$s$)(100$ mK$)$&1.5&5.1&0.8&14&1.8\\
meas $\tau_\phi(\mu$s$)(T/$mK$)$&---&---&0.02(25)&0.01(25)&0.50(15)\\
calc $\Omega \tau_\phi/2\pi(100$ mK$)$&5100&3000&2700&97000&30000\\
meas $\Omega \tau_\phi/2\pi(T/$mK$)$&---&---&68(25)&69(25)&8000(15)\\
\hline
\end{tabular}
\end{table*}
\par

\section {CONCLUSIONS}

Despite ongoing studies over more than two decades, the origin of
$1/f$ noise in the critical current of Josephson junctions is still
not fully understood.  Although there is strong evidence that the
noise derives from a superposition of random telegraph signals
produced by charge trapping and untrapping processes, the origin of
the $T^2$ dependence observed by Wellstood \cite{Wellstood-thesis}
remains puzzling.  This temperature dependence can be explained within
the framework of a two-well potential in which the two barrier heights
are independent random variables, provided one assumes
thermally-activated processes rather than the tunneling processes one
might expect.  Furthermore, the absence of a temperature dependence of
the form $\exp(-\Delta/k_B T)$ at low temperatures is difficult to
understand in a picture in which the trap exchanges single electrons
with superconducting electrodes.  Clearly more work is required to
understand this behavior. We found that the measured spectral density
of the $1/f$ noise in the critical current of junctions with different
materials and a wide range of areas and critical currents scales
surprisingly well as $[144(I_0/\mu$A$)^2/(A/\mu$m$^2)]($pA$)^2/$Hz at
$4.2$ K. Based solely on the results of Wellstood we have chosen to scale
this number with $(T / 4.2$ K$)^2$ to predict the $1/f$ noise at
$100$ mK.  How well this scaling remains valid as more junctions are
investigated and whether the $T^2$ dependence holds down to (say) $10$ mK are questions that should be addressed with some urgency.  These
measurements must of necessity be made with a SQUID amplifier; the use
of submicron junctions with relatively high critical currents should
enhance the magnitude of the noise and make its observation more
straightforward.

For four different qubits we calculated the parametric effect of small
changes in the critical current $I_0$ on the energy separation $\hbar
\Omega$ at the operating point.  Using the normalized parameter
$\Lambda=|I_0 d \Omega / \Omega d I_0|$ and the extrapolated magnitude
of the $1/f$ noise we investigate dephasing in these qubits at $0.1$ K. In agreement with the treatment of Martinis {\it et al.},
\cite{Martinis-noise}  we find that the sources of decoherence
accumulate as $t^2$, so that the decoherence is not interpretable as a
rate.  Rather, the frequency is different each time a measurement is
made.  In all cases where $\tau_{\phi}$ has been measured, the
calculated values due to critical current $1/f$ noise are
%one or two orders of magnitude
greater than the measured values.  Furthermore, if the $T^2$
dependence of the $1/f$ noise does continue at temperatures down to
(say) $10$ mK, the predicted decoherence time, which scales as $1/T$,
will become an order of magnitude longer at this temperature.
Nonetheless, although critical current $1/f$ noise appears not to be
the limiting source of decoherence in experiments conducted to date,
ultimately this mechanism will present an upper bound on
$\tau_{\phi}$.

Although the level of $1/f$ noise is remarkably constant for existing
junction technologies, there may be alternative schemes for growing
the tunnel barrier which reduce the number of charge traps in the
barrier, and hence reduce the noise. We note also that even in the
presence of low frequency noise, the use of various pulse sequences,
such as spin echoes, \cite{Nakamura2,Quantronium,Chiorescu,Hahn}  or
bang-bang pulses \cite{bang-bang} can significantly reduce its
effects.

Finally, in the case of flux qubits this formulation could be extended
to the effects of $1/f$ flux noise originating from either magnetic
vortex motion or current noise in the current supply by calculating
the quantity $d \Omega /d \Phi$.

\medskip

\begin{acknowledgments}
  We thank Tony Leggett, John Martinis, Michael Weissman, Fred
  Wellstood, and Frank Wilhelm for useful discussions.  This work was
  supported in part by the Air Force Office of Scientific Research under
  Grant F49-620-02-1-0295, the Army Research Office under Grant
  DAAD-19-02-1-0187, and the National Science Foundation under Grants
  EIA-020-5641 and EIA-01-21568.  DVH thanks the Miller Institute at the
  University of California, Berkeley, for Fellowship support.
\end{acknowledgments}

\bibliography{1-f-paper}

\end{document}